\documentstyle[sprocl,epsfig]{article}

\def\be{\begin{equation}}
\def\ee{\end{equation}}
\def\bea{\begin{eqnarray}}
\def\eea{\end{eqnarray}}
\def\md{M_D}
\def\bit{\begin{itemize}}
\def\eit{\end{itemize}}

\def\rts{\sqrt s}

\def\fbi{~{\rm fb}^{-1}}

\def\slasht#1{#1\hskip-8pt/\hskip5pt}
\def\mpl{\overline M_{Pl}}
\def\to{\rightarrow}

\def\del{\delta}
\def\eps{\epsilon}

    \def\fillboxx#1#2{\hbox to #1{\vbox to #2{\vfil}\hfil}   }

\def\gev{~{\rm GeV}}

\def\gam{\gamma}

\def\mh{m_h}

\def\etmiss{\slasht E_T}

\begin{document}

\title{
INVISIBLE HIGGS IN LARGE EXTRA DIMENSION MODELS  }

\author{DANIELE DOMINICI}

\address{Dipartimento di Fisica, Universit\`a di Firenze,  I-50019 Sesto
F., Italy\\
I.N.F.N.,
Sezione di Firenze,  I-50019 Sesto F., Italy\\}


\maketitle\abstracts{
In large extra dimension models the presence of an interaction between the Ricci scalar curvature and the Higgs doublet of the Standard Model 
can give rise to an invisible decay of the Higgs to Kaluza Klein graviscalars.
The corresponding invisible width can cause a suppression of the LHC rates 
of a light Higgs in the visible channels below 5$\sigma$
in  some regions of the parameter space of the model. However in  such  regions
the Higgs can be discovered through its invisible decay.
The combination of the  measurements done at the LHC and the  LC can determine to some accuracy the parameters of the model.
}

\section{Introduction}
In large extra dimension  models the interaction between
the Higgs  doublet field $H$ and the 
Ricci scalar curvature $R$ of the induced 4-dimensional metric
$g_{ind}$,
given by 
$
S=-\xi \int d^4 x \sqrt{g_{ind}}R(g_{ind})H^\dagger H\,,
$
after the usual shift $H=({v+ h\over \sqrt{2}},0)$,
 leads to the following mixing term \cite{Giudice:2000av}
\begin{equation}
{\cal L}_{\rm mix}=\epsilon  h \sum_{\vec n >0}s_{\vec n}
\label{mixing}
\end{equation}
with
\be
\eps=-{2\sqrt 2\over \mpl}\xi v \mh^2\sqrt{{3(\del-1)\over \del+2}}\,.
\label{epsdef}
\ee
Above, $\mpl=(8\pi G_N)^{-1/2}$ is the reduced Planck mass, $\delta$ is the
number of extra 
dimensions, $\xi$ is a dimensionless parameter and
$s_{\vec n}$ is a graviscalar KK excitation with
mass $m_{\vec n}=2\pi\vert  \vec n\vert /L$, $L$ being the
size of each of the extra dimensions.
This mixing generates an oscillation of the Higgs itself into the closest
Kaluza Klein 
graviscalar levels and therefore an invisible decay for the Higgs.
The corresponding width
can be calculated by extracting the imaginary part of the Higgs self energy contribution
due to the mixing  in eq. (\ref{mixing}) \cite{Giudice:2000av,Wells:2002gq}
(see also \cite{Allanach:2004ub}) and is given by
\bea
\Gamma_{h\to inv}&=&
{\pi\over 2} m_h^{\del -3}\eps^2 {\mpl^2\over M_D^{2+\del}}
{\pi^{\del/2}\over \Gamma(\del/2)}\,
\nonumber \\
&\sim& (16\,MeV) 20^{2-\delta } \xi^2
S_{\delta-1}\frac {3(\delta -1)}
{\delta +2} \left ( \frac {m_h}{150\, GeV} \right )^{1+\delta}
\left ( \frac 
{3\, TeV} {M_D}\right )^{2+\delta}\,.
\label{invwidth}
\eea

 \begin{figure}[htb]
     \begin{center}
     \begin{tabular}{cc}
     \mbox{\epsfig{file=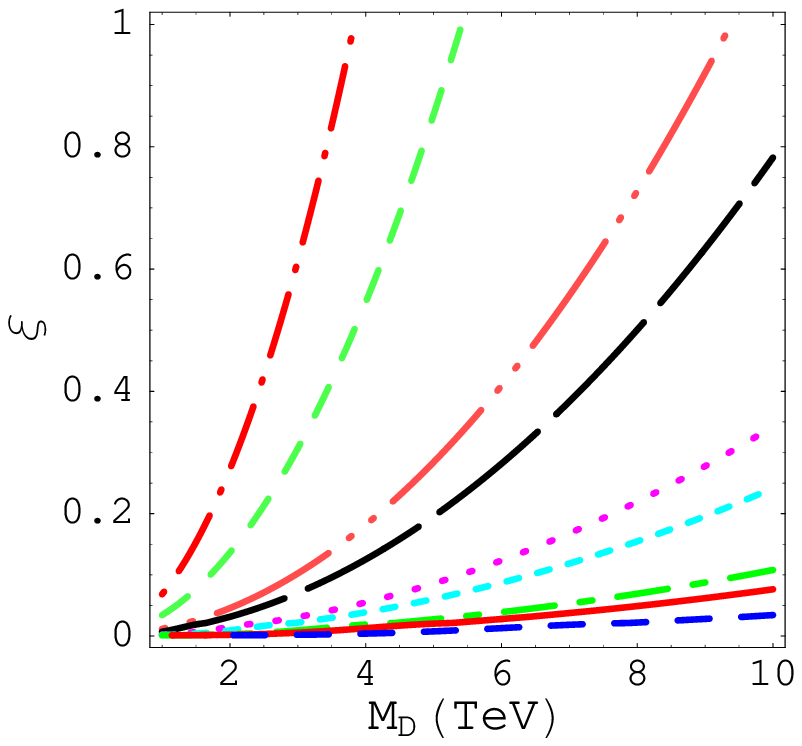,width=5.7cm}}&
     \mbox{\epsfig{file=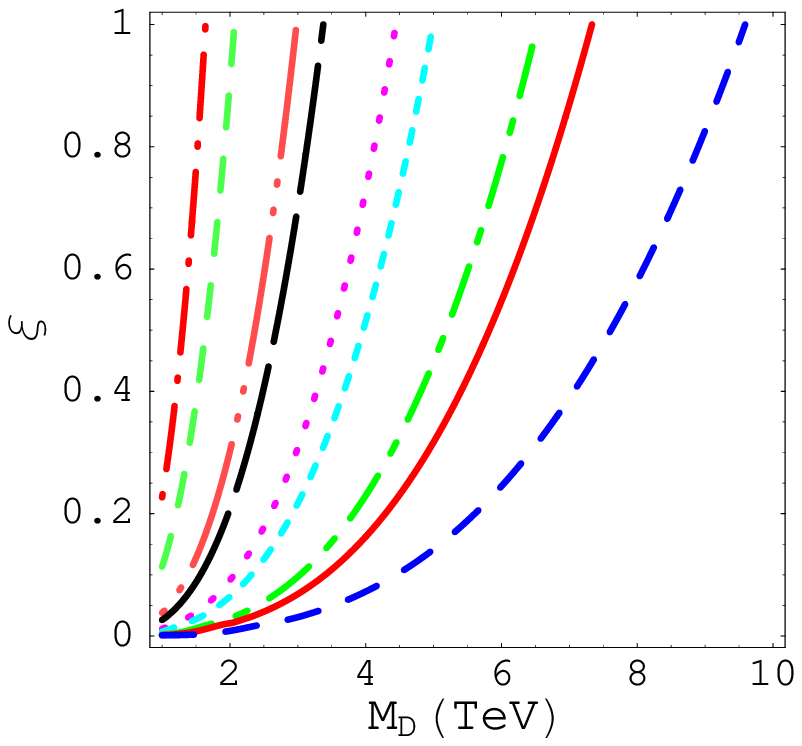 ,width=5.7cm}}\\
\mbox{\epsfig{file=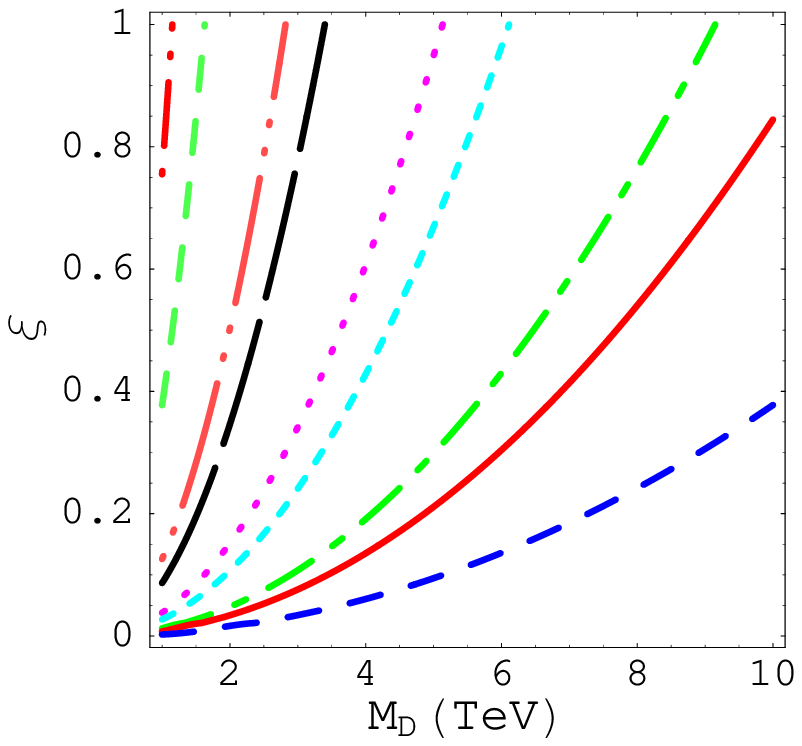,width=5.7cm}}&
     \mbox{\epsfig{file=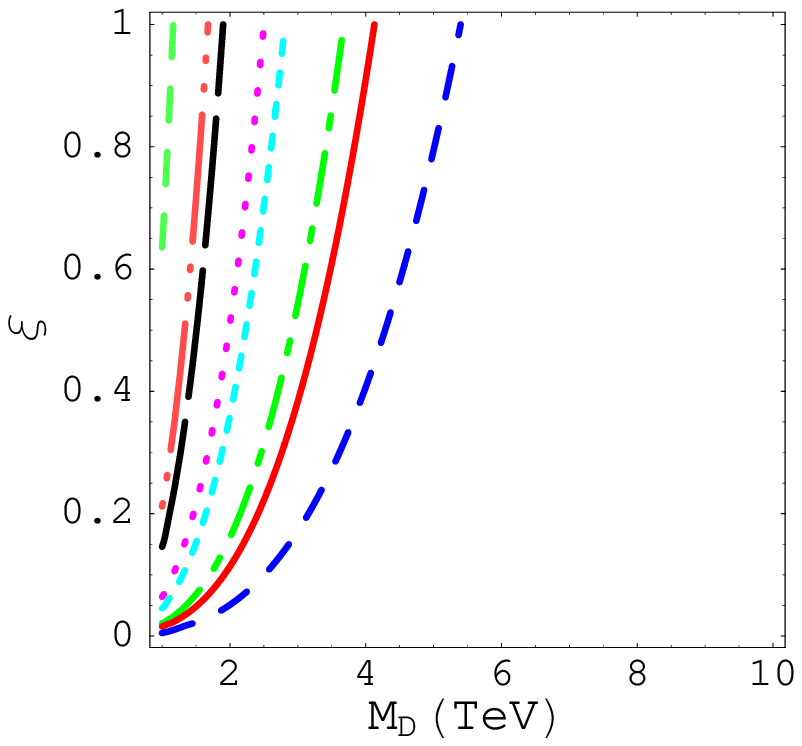 ,width=5.7cm}}
     \end{tabular}
     \end{center}
     \caption{Contours of fixed $BR(h\to inv)$  in the
$M_D$(TeV) -- $\xi$ parameter space for  $\del=2$ (left)
and $\del=4$ (right) in the upper part for   $m_h=120\gev$ and
in the lower part for $m_h=237\gev$. 
 In order of increasing $\xi$ values, the 
contours correspond to:
 $0.0001$
(large blue dashes), $0.0005$ (solid red line), $0.001$  (green
long dash -- short dash line), $0.005$ (short cyan dashes), 
$0.01$ (purple dots), $0.05$ (long black dashes), $0.1$ (chartreuse
long dashes with
double dots), $0.5$ (green dashes), and $0.85$ (red long dash,
short dot line at high $\xi$ and
low $M_D
$)
}
     \label{fig:br}
     \end{figure}


\section{Detecting the Higgs at the LHC and LC}

\begin{figure}[ht]
     \begin{center}
     \begin{tabular}{cc}
     \mbox{\epsfig{file=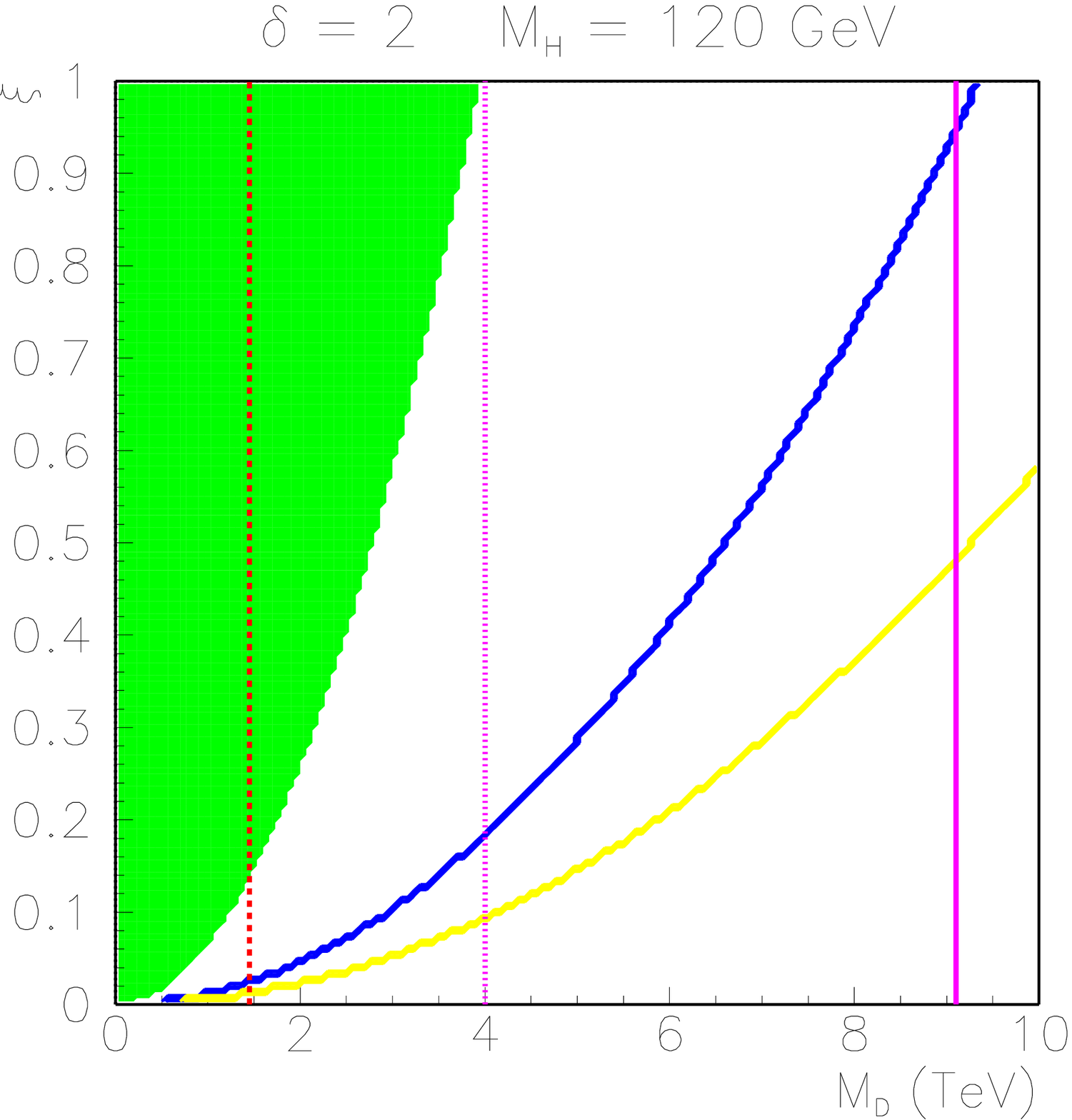,width=5.7cm}}&
     \mbox{\epsfig{file=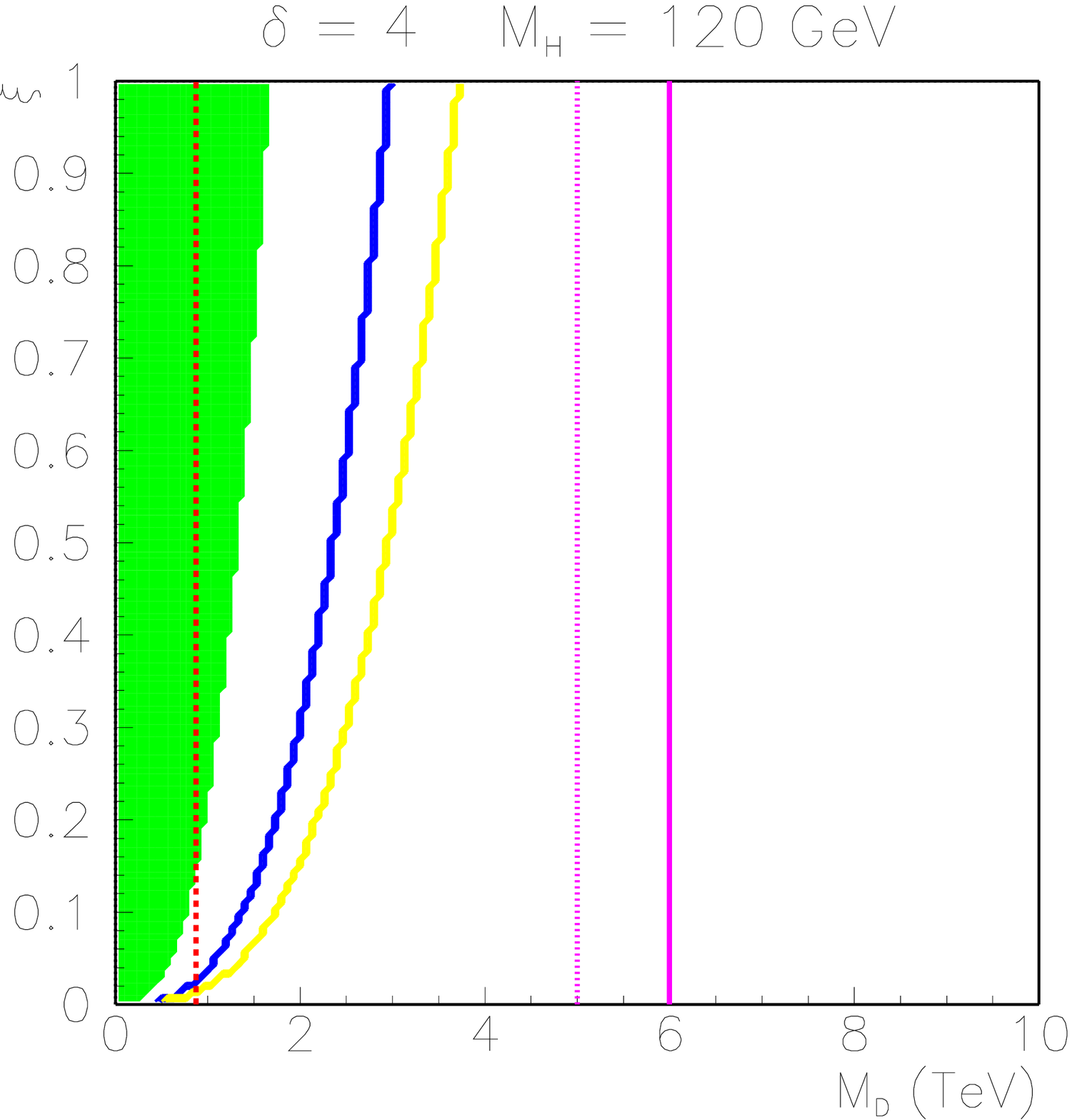 ,width=5.7cm}}
     \end{tabular}
     \end{center}
     \caption{Invisible decay width effects 
for 
 $m_h$ = 120~GeV,  $\delta$: 2
  (left), 4 (right). The
 green (grey) regions indicate where the Higgs signal at the LHC
  drops below the 5~$\sigma$ threshold for 100 fb$^{-1}$.  The
 regions above the blue (bold) line are  where the LHC invisible Higgs signal in the $WW$-fusion
  channel exceeds 5~$\sigma$ significance. The solid vertical line shows the upper limit on
 $M_D$ which can be probed at the $5~\sigma$ level
 by the analysis of jets/$\gamma$ with
 missing energy at the LHC.  The middle dotted vertical line 
  shows the value of $M_D$ below which the
  theoretical computation at the LHC is ambiguous.  The dashed vertical line at the lowest $M_D$ value is
  the 95\% CL lower limit coming from  Tevatron and LEP/LEP2.
  The regions above the yellow (light grey) line 
are where the LC invisible Higgs signal will exceed
$5~\sigma$.
}
     \label{fig:120-2-4}
     \end{figure}

Fig.~\ref{fig:br}  shows
that the branching ratio of the Higgs into invisible states
can be substantial  for $M_D$ values in the  TeV range both when
   $m_h=120$ GeV (upper part), therefore below the $WW$ threshold,  and 
 when $m_h=237$ GeV (lower part), a value  greater than the $WW$ threshold 
and corresponding to the 95\% CL limit from LEP data with $m_t=178$ GeV.
As a consequence 
this invisible
width causes a significant suppression of the LHC Higgs rate in the
standard visible channels and
for any given value of the Higgs boson mass, there is a considerable
parameter space region where the invisible decay width of the Higgs boson
could be the first measured phenomenological effect from extra
dimensions. This is exemplified in Fig.~\ref{fig:120-2-4}
for $m_h$ = 120~GeV,  $\delta$= 2
  (left), 4 (right).
In the green (light grey) region the Higgs signal in
standard channels drops below the 5~$\sigma$ threshold with 100
fb$^{-1}$ of LHC data. But in the area above the bold blue line the
LHC search for invisible decays in the fusion channel yields a signal
with an estimated significance exceeding 5~$\sigma$. In conclusion, 
whenever the Higgs boson sensitivity is lost due to
the suppression of the canonical decay modes, the invisible rate is
large enough to still ensure detection through the $WW$ fusion channel. 

\begin{figure}[htb]
     \begin{center}
     \begin{tabular}{cc}
     \mbox{\epsfig{file=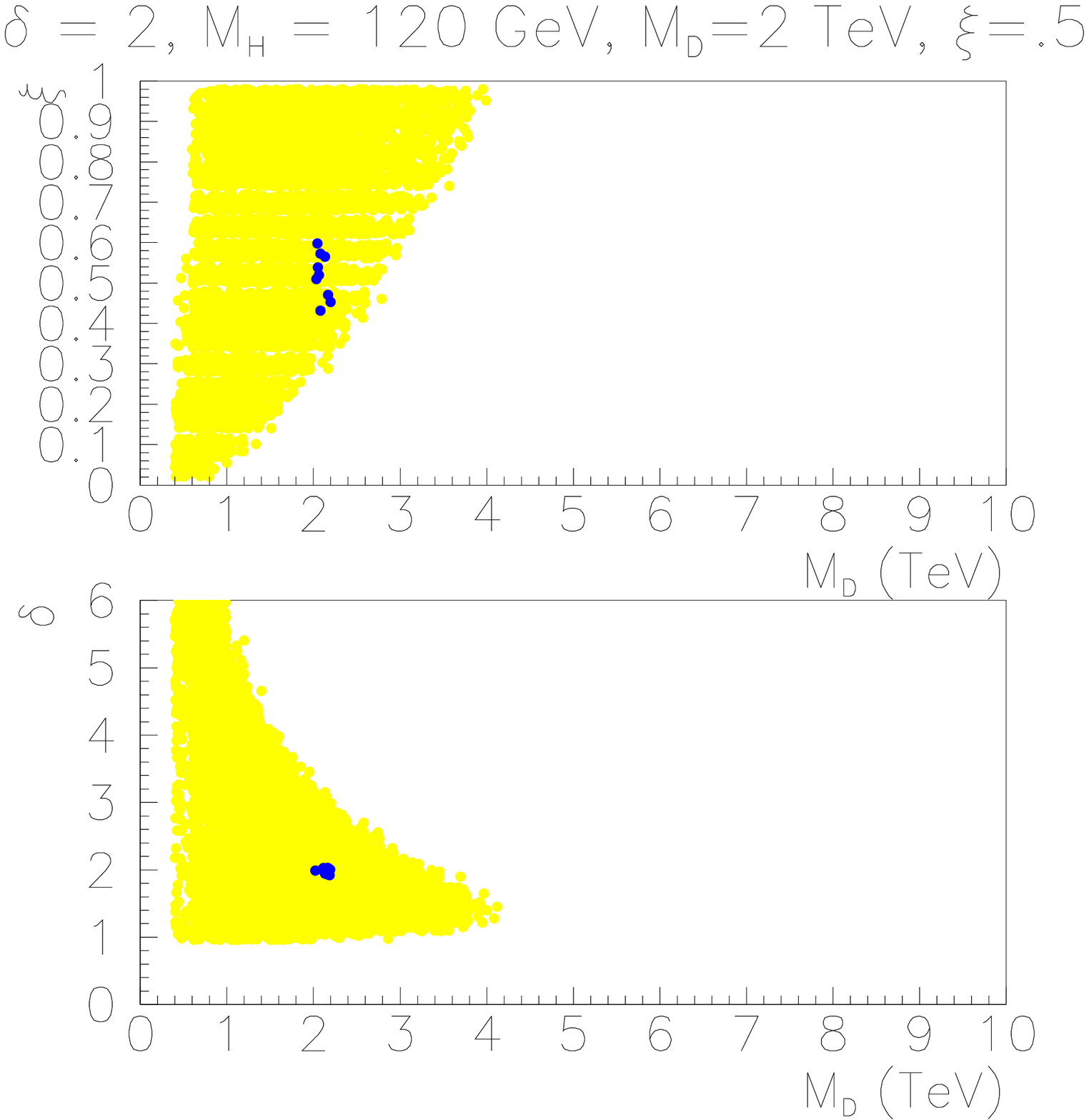,width=5.7cm}}&
     \mbox{\epsfig{file=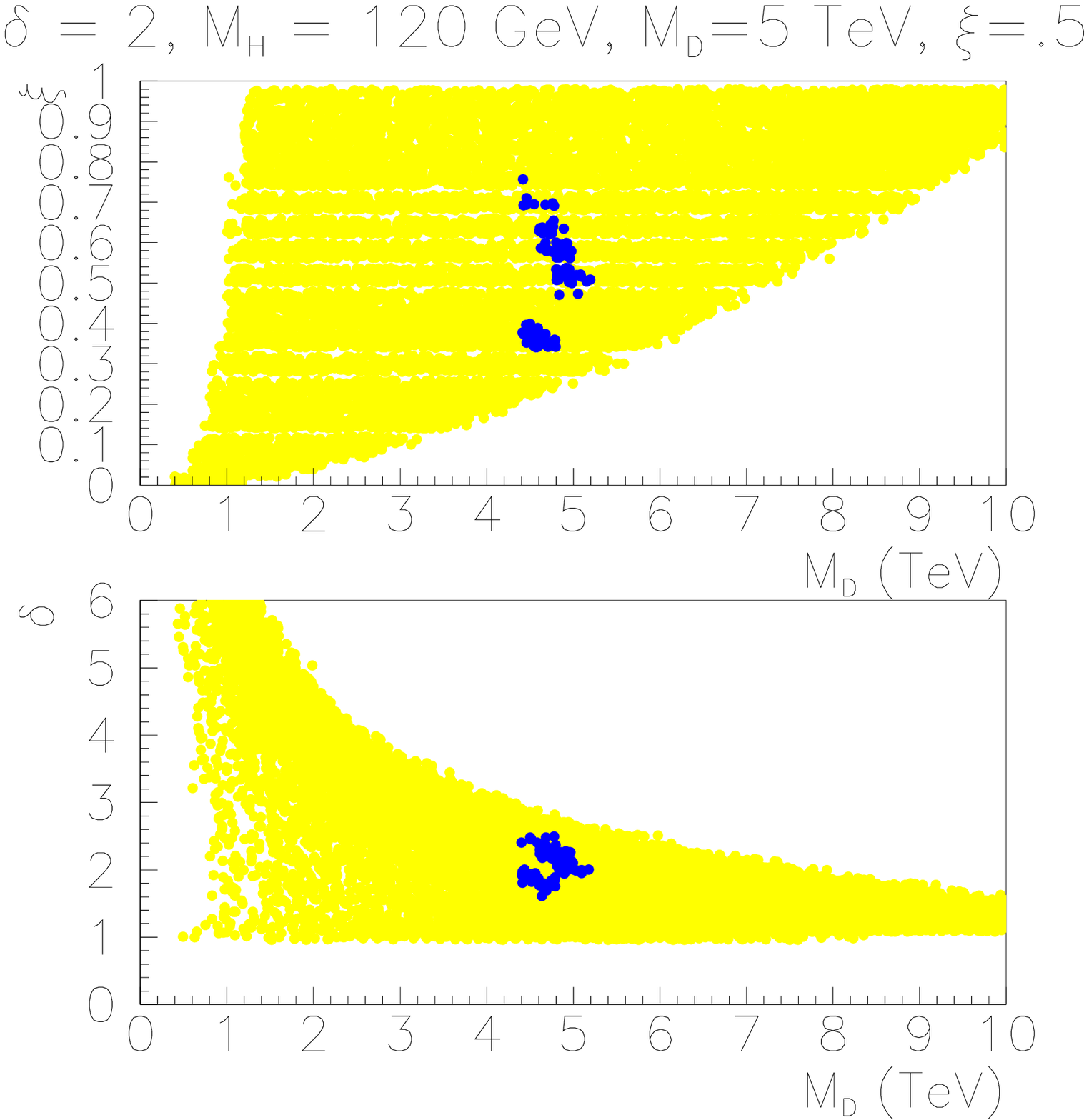 ,width=5.7cm}}
     \end{tabular}
     \end{center}
     \caption{95\% CL contours for the determination of the ADD parameters,
  assuming $m_{h}=120\gev$.
The larger light grey (yellow) regions are the 95\% CL regions 
 using only $\Delta\chi^2(LHC)$,
the smaller dark grey (blue) regions  using $\Delta\chi^2(LHC+LC)$.
}
     \label{fig:prmtrs}
     \end{figure}

Fig.~\ref{fig:prmtrs}  shows 95\% CL contours for determination of the ADD parameters,
  $\md$, $\xi$ and $\delta$ assuming $m_{h}=120\gev$ obtained by combining
LHC and LC  signal in visible and invisible channels and LC $\gamma+\etmiss$ 
 cross section measurements at two different energies.
The plots
  are all obtained for $\delta=2$ and $\xi=0.5$,
assuming $L=100\fbi$ at the LHC, $\rts=350\gev$ Higgs measurements
at the LC,  $\rts=500\gev$ and $\rts=1000\gev$ $\gam+\etmiss$
measurements at the LC with $L=1000\fbi$ and $L=2000\fbi$ at the
two respective energies.

In conclusion the accuracy of the determination of the three parameters of the model is relatively good unless both $\delta$ and $M_D$ are large or $\xi<<1$.

\begin{center}
{\bf Acknowledgments}
\end{center}
I would like to thank M. Battaglia, J. Gunion and J. Wells for their
collaboration on the topics discussed in this talk.



\end{document}